\documentclass[amsmath,twocolumn,amssymb,floatfix,superscriptaddress]{revtex4-1}

\usepackage{graphicx}
\usepackage{amssymb}
\usepackage{amsmath}
\usepackage{color}
\usepackage{ wasysym }
\usepackage{hyperref}
 \usepackage{tabu}

\def\barray{\begin{array}}
\def\earray{\end{array}}
\def\be{\begin{equation}}
\def\ee{\end{equation}}
\def\ben{\begin{equation} \nonumber}
\def\een{\end{equation}}
\def\ban{\begin{eqnarray*}}
\def\ean{\end{eqnarray*}}
\def\ba{\begin{eqnarray}}
\def\ea{\end{eqnarray}}

\def\({\left(}
\def\){\right)}

\graphicspath{{./fig/}}

\begin{document}

\title{Information gains from Monte Carlo Markov Chains}
\author{Ahmad Mehrabi}
\author{A. Ahmadi }
\affiliation{Department of Physics, Bu-Ali Sina University, Hamedan, Iran}


\begin{abstract}
In this paper, we present a novel method for computing the relative entropy as well as the expected relative entropy using an MCMC chain.  The relative entropy from information theory can be used to quantify differences in posterior distributions of a pair of experiments. In cosmology, the relative entropy has been proposed as an interesting tool for model selection, experiment design, forecasting and measuring information gain from subsequent experiments. In contrast to Gaussian distributions, these quantities are not generally available analytically and one needs to use numerical methods to estimate them which are certainly computationally expensive. We propose a method and provide its python package to estimate the relative entropy as well as expected relative entropy from a posterior sample.  We consider the linear Gaussian model to check the accuracy of our code. Our results indicate that the relative error is below $0.2\%$ for sample size larger than $10^5$ in the linear Gaussian model. In addition, we study the robustness of our code in estimating the expected relative entropy in this model.         
\end{abstract}
\maketitle

\section{Introduction}
In contrast to a few decades ago, there are a large number of probes in cosmology, which provide us remarkable information about content and evolution of the Universe. These data sets have been extensively used to study and constrain free model parameters in literature (see \cite{Mehrabi:2015hva,Mehrabi:2016exz,Rezaei:2017yyj,Mehrabi:2018dru}and references therein). Bayesian inference provides a common and widely used method to constrain free model parameters. In this case, we  update a prior probability density in parameter space to obtain posterior distribution using an observational data. Since an analytic solution in the Bayesian inference is very limited, one has to develop a numerical method to find the posterior. Among these, the Monte Carlo Markov Chain (MCMC) techniques are widely accepted and used in different problems. The purpose of an MCMC algorithm is to construct a sample of points in parameter space which is called a chain and then obtain posterior probability density from it. The simplest and widely used MCMC algorithm is Metropolis-Hasting \cite{Hasting:1970} but considering different situations other algorithms like Gibbs sampling \cite{gibs-sam:2009,gibs-sam:2011} and Hamilton Monte-Carlo \cite{hami-sam:2011}  have been used to obtain the posterior distributions. To quantify the difference between probability distributions from different surveys, a robust framework is needed.

Initially motivated from information theory, the relative entropy or Kullback-Leibler divergence has been proposed to measure differences in two probability densities \cite{kull:2011} . In addition, this method has been used for experiment design and forecasting \cite{Farhang:2011ud,Paykari:2012ne,Amara:2013swa} as well as model selection \cite{Kunz:2006mc,Verde:2013wza} in cosmology. Moreover, the relative entropy has been introduced as a new tool to measure information gain from successive experiment \cite{Seehars:2014ora,Grandis:2015qaa} as well as a tool to measuring tensions among datasets within a given model \cite{Seehars:2015qza,Nicola:2018rcd}. The relative entropy quantifies both statistical precision and shifts of confidence regions and by disentangling these contributions, one can measure change of confidence regions and shifts of parameters from two different datasets.  In the limit of Gaussian distributions, the relative entropy has an analytic solution but in a general case, one should use numerical method to obtain it. Since in most cases the probability distributions, coming from a MCMC chain, it would be a remarkable task to provide the relative entropy from an MCMC chain. In this work, we introduce a method and provide a python package to estimate the relative entropy from an MCMC chain.    

Given two datasets $D_1$ and $D_2$, it is straightforward to find the posterior probability distributions and then the relative entropy between these two distributions.
As we mentioned above, the relative entropy consist of two contributions, information gain in precision and shifts in parameter space. To distinguish these two contributions, one can use the constrains from $D_1$ dataset to anticipate the expected relative entropy for $D_2$ dataset by assuming both datasets are described by the same model. The difference between the relative entropy and the expected one is called surprise and has been introduced in \cite{Seehars:2014ora} as a remarkable tool to measure consistency between datasets in a given model. The expected relative entropy has an analytic solution in the case of two Gaussian distributions but for a general case, we need to use a numerical method to estimate it. To do this, many algorithms have been proposed  in literature \cite{Drovandi:2013,Long:2013,Huan:2013}. In this work, we introduce a python package to estimate the expected relative entropy base on the algorithm proposed in  \cite{Huan:2013}.
 
This work is organized as follows. In section \ref{sect:rel_ent} we review the formalism of  relative entropy and present  it for two  Gaussian distributions. In section \ref{sect:surprise}, we argue about the surprise and its close-form in the Gaussian limit. In section \ref{sec:linear}, we present the linear Gaussian model and compare the exact results from those of our code to check its accuracy.  Finally, in \ref{conclude}, we conclude and highlight importance of our method.

\section{Information gain base on the relative entropy }\label{sect:rel_ent}
The likelihood is the probability of the data $\mathcal{D}$ given the value of the parameters  and is a crucial quantity in parameter inference. Given a likelihood, it is straightforward to update a prior information on parameters $\pi(\Theta)$ to obtain the posterior $P(\Theta|\mathcal{D})$ through Bayes' theorem:
 \begin{equation}\label{eq:bays}
 P(\Theta|\mathcal{D}) = \frac{\mathcal{L}(\Theta;\mathcal{D})\pi(\Theta)}{E(\mathcal{D})}\; ,
 \end{equation}
where $\mathcal{L}(\Theta;\mathcal{D})$ is the likelihood function for the data and the denominator is the Bayesian evidence which is given by
 \begin{equation}\label{eq:evi}
E(\mathcal{D})=\int d\Theta \mathcal{L}(\Theta;\mathcal{D})\pi(\Theta).
 \end{equation}
 In this process, one can measure information gain from updating the prior to posterior  via the relative entropy or Kullback-Leibler. The relative entropy between two probability distributions $P_1(\Theta)$ and $P_2(\Theta)$ is given by:
\begin{equation}\label{eq:rel-ent}
D(P_2||P_1) \equiv \int d\Theta\, P_2(\Theta) \log \frac {P_2(\Theta)} {P_1(\Theta)}.
\end{equation}
The relative entropy is always positive and equals to zero only for $P_1(\Theta)=P_2(\Theta)$. Apart from not being symmetric in $P_1$ and $P_2$, the relative entropy is invariant under invertible transformations in $\Theta$. 

For two Gaussian distributions $P_1(\Theta)= \mathcal{N}(\Theta;\Theta_1,\Sigma_1)$ and $P_2(\Theta)= \mathcal{N}(\Theta;\Theta_2,\Sigma_2)$ the relative entropy is given by:
\begin{equation}\label{eq:rel-gus}
\begin{aligned}
&D(P_2||P_1) = \frac 1 2 (\Theta_1 - \Theta_2)^T \Sigma_1^{-1} (\Theta_1 - \Theta_2)\\
&\qquad + \frac 1 2 \left(\text{tr}(\Sigma_2\Sigma_1^{-1}) -d - \log \frac {\det(\Sigma_2)} {\det(\Sigma_1)} \right).
\end{aligned}
\end{equation}
The first term measure the significance of mean shift and the second term quantifies change in precision. In a general case, the probability distributions are not Gaussian so developing a method to estimate the relative entropy for any arbitrary distributions, is a remarkable task. Assuming $P_2(\Theta)$ as a posterior and using Eq.(\ref{eq:bays}), the relative entropy can be rewritten as:
\begin{equation}\label{eq:rel-ent-arb}
D(P_2||P_1) =  -\log E(\mathcal{D}) +\int d\Theta P_2(\Theta)\log\mathcal{L}(\Theta)\;.
\end{equation}     
Given a sample of posterior, the second integral can be easily estimated from $<\log\mathcal{L}(\Theta)>_{P_2}$ so knowing the evidence ,one can estimate the relative entropy from a sample for any arbitrary distributions. We provide  a python package (available at https://github.com/ahmadiphy/MCKLdivergence) to estimate the relative entropy using Eq.(\ref{eq:rel-ent-arb}). Inputs are a sample chain and $\log (\rm{prior})$ at each sample in the chain. The code estimate the first term using method introduced in \cite{Heavens:2017afc} using kth nearest-neighbour distances in parameter space.

\section{Expected relative entropy and surprise}\label{sect:surprise}
Considering a prior and likelihood function, it is possible to find several realizations of data. Assuming $P(\mathcal{D}^{\prime}|\mathcal{D})$ as the probability of obtaining $\mathcal{D}^{\prime}$ given $\mathcal{D}$, the expected relative entropy is given by
  \begin{equation}\label{eq:exp-rel}
 <D> = \int d\mathcal{D}^{\prime } P(\mathcal{D}^{\prime}|\mathcal{D})     D(P_2||P_1)\; ,
 \end{equation}
 where $P(\mathcal{D}^{\prime}|\mathcal{D})$ is given by:
   \begin{equation}\label{eq:pddp}
 P(\mathcal{D}^{\prime}|\mathcal{D}) = \int d\Theta P(\Theta|\mathcal{D})P(\mathcal{D}^{\prime}|\Theta)\; .
 \end{equation}
 Notice that, it is possible to consider the prior from one dataset for example $\mathcal{D}_1$ and likelihood from $\mathcal{D}_2$ so the expected relative entropy can be estimated between two datasets. The surprise is defined via \cite{Seehars:2014ora}
\begin{equation}\label{eq:sur}
 S = D - <D>,
  \end{equation} 
which scatters around zero. A positive value of S indicates that posterior is more different that what we expect and a negative value means the constrains are more consistent than expected a priori.  

It has been proved that S follows a generalized $\chi^2$ distribution \cite{Seehars:2014ora} for Gaussian distributions and given a particular value of S, one can measure the probability for measuring S that deviates from zero by more than S. This quantity is the so called p-value for hypothesis that both datasets are consistent within the considered model and a small p-value indicates evidence against the hypothesis. 

In the limit  of two Gaussian posteriors, the surprise is given by 
  \begin{equation}\label{eq:sur-gus}
  S = \frac 1 2 \{(\Theta_1 - \Theta_2)^T \Sigma_1^{-1} (\Theta_1 - \Theta_2) - \rm tr(1\pm \Sigma_2\Sigma_1^{-1})\},
  \end{equation} 
 where the - holds when posterior of $D_1$ is used as a prior to obtain posterior of $D_2$ and the + holds when a wide prior is used for both posteriors. Since in  a general case, posteriors derived from $D_1$ and $D_2$ are not Gaussian, we need a general approach for obtaining surprise. 
 
 In the Bayesian experimental design, the expected relative entropy is a well known quantity. In fact many algorithms have been proposed to obtain this quantity in a general non-linear cases \cite{Drovandi:2013,Long:2013,Huan:2013}. Among these, a simple method has been proposed in \cite{Huan:2013}, where the expected relative entropy can be estimated from 
 \begin{equation}\label{eq:exp-rel-num1}
 <D>\approx\frac 1 n \sum_{i=1}^{n}\{ \log P(\mathcal{D}_2^{i}|\Theta_i) -\log P(\mathcal{D}_2^{i}|\mathcal{D}_1)\},
 \end{equation}       
 where the second term can be estimated from
  \begin{equation}\label{eq:exp-rel-num2}
 P(\mathcal{D}_2^{i}|\mathcal{D}_1)\approx \frac 1 n \sum_{j=1}^{n} P(\mathcal{D}_2^{i}|\Theta_j).
 \end{equation}    
 In the above formula, $\Theta_i$ is a sample from $\mathcal{D}_1$ posterior and $\mathcal{D}_2^{i}$ is a sample of simulated data from $\mathcal{D}_2$ likelihood. Given a sample of $\Theta$ from $\mathcal{D}_1$ posterior and the likelihood function of $\mathcal{D}_2$, it is possible to estimate the expected relative entropy and then the surprise from the above formula. We provide a class in our python package to estimate the expected relative entropy using above algorithm.  In this case, inputs are a sample of $\Theta$ from $\mathcal{D}_1$ posterior, the covariance of the likelihood $(\Sigma)$, the model function $F(\Theta)$, number of sample $(l)$ to be used to estimate expected relative entropy and the dimension of the data $(n)$. The code uses $l$ given sample value to simulate $l$ data from a Gaussian likelihood
 \begin{equation}
 \mathcal{L}(\Theta;\mathcal{D}) \sim e^{(\mathcal{D}-F(\Theta))^{T}\Sigma^{-1}(\mathcal{D}-F(\Theta))},
 \end{equation}  
 and then uses them to estimate the expected relative entropy from Eq.(\ref{eq:exp-rel-num1}). Notice that the user defined model function should return a vector of dimension $n$. The current version of the code adopt only a Gaussian likelihood and we plan to update it to a general case in subsequent updates.  
 
\begin{figure}
 \centering
  \includegraphics[width=.5\textwidth]{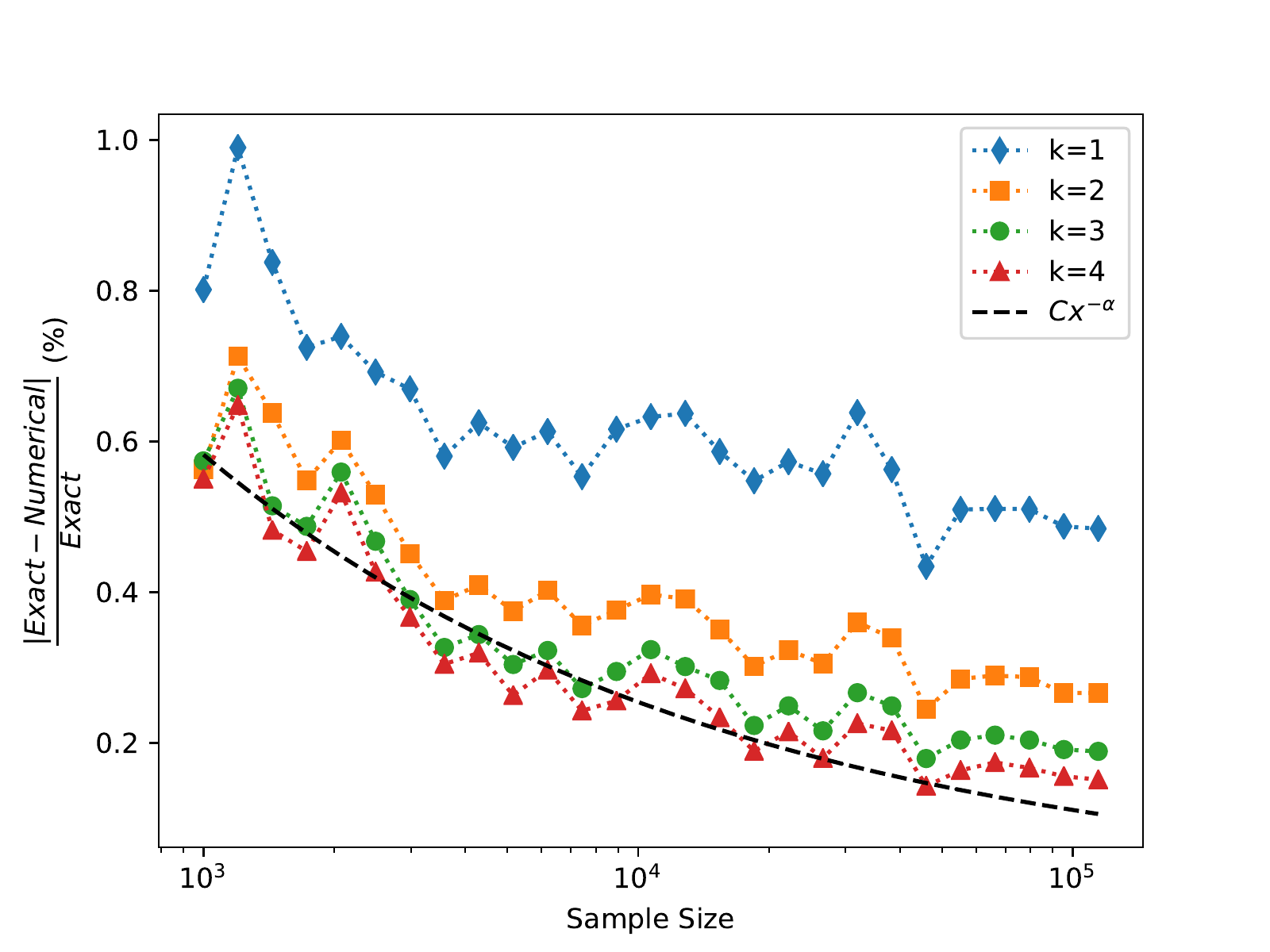}
  	\caption{Relative difference of the exact and estimated relative entropy in 3D case.}
  	\label{fig:rel-ent-3d}
  \end{figure}
  
  \begin{figure}
  	\centering
  	\includegraphics[width=.5\textwidth]{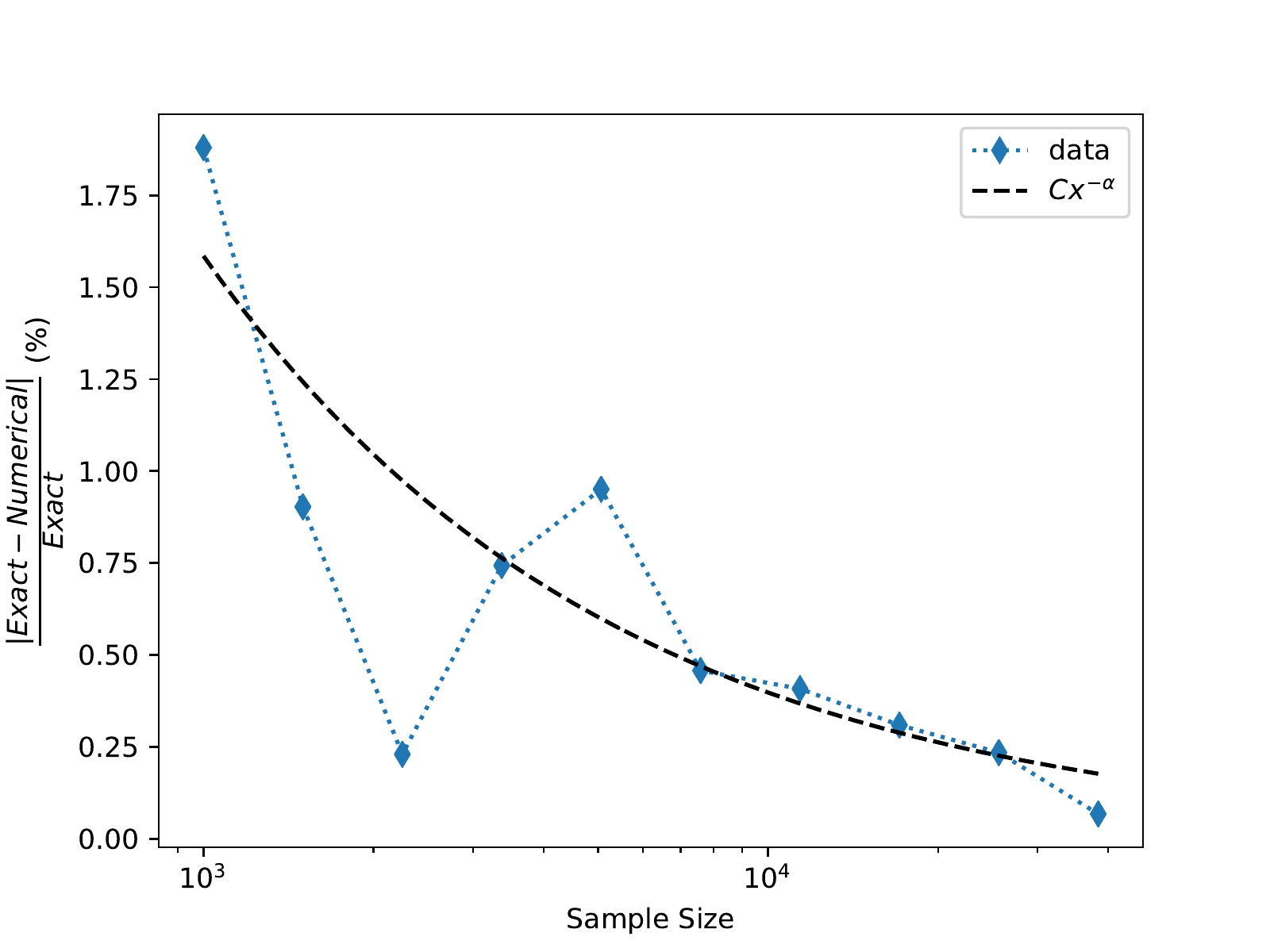}
  	\caption{Relative difference of the exact and estimated expected relative entropy in 3D case.}
  	\label{fig:rel-eent-3d}
  \end{figure}
\section{Linear Gaussian model and accuracy of the method}\label{sec:linear}
In order to check the accuracy of our code, we consider the linear Gaussian model which has an analytic solution for both relative entropy and expected relative entropy.  We consider a Gaussian prior $\mathcal{N}(\Theta;\Theta_1,\Sigma_1)$ on $\Theta$ and a Gaussian likelihood $\mathcal{N}(\mathcal{D};F(\Theta),\mathcal{C})$ on the data. The model function $F(\Theta)$ must be linear in $\Theta$, we assume 
\begin{equation}
F(\Theta) = F_0 + M\Theta,
\end{equation}
where $M_{ij} = X^{j}(x_i)$ is a matrix evaluated at some arbitrary points $x_i$ and $F_0$ is a constant vector. In this formalism, $X(x)$ are the basic functions
 which can very well be a non-linear function of $x$. Notice that the model function must be linear in $\Theta$, not necessarily in the basic functions.  
 Using Bayes' theorem, the posterior is a Gaussian with the following covariance and mean
\begin{equation}\label{eq:pos_lin}
\begin{aligned}
&\Sigma_2 = (\Sigma_1^{-1}+M^{T}\mathcal{C}^{-1}M)^{-1}\\
&\Theta_2 = \Sigma_2(\Sigma_1^{-1}\Theta_1+M^{T}\mathcal{C}^{-1}(\mathcal{D}-F_0) )
\end{aligned}
\end{equation}      
Having both Gaussian prior and posterior, we can use Eq.(\ref{eq:rel-ent}) and Eq.(\ref{eq:exp-rel}) to compute the relative and expected relative entropy. After this, we generate a sample chain using Hamilton algorithm and compute these quantities by our code. The results in 3 dimensions have been shown in Figs (\ref{fig:rel-ent-3d} and \ref{fig:rel-eent-3d}).In the case of relative entropy, for $k=1$ and with $10^4$ samples, the relative error is around $0.6\%$ but it decreases for a larger k and is below $0.2\%$ for $k=4$ with $10^5$ samples. As we show in Fig (\ref{fig:rel-ent-3d}), the relative entropy decreases like a power law by increasing size of the sample. 
The expected relative entropy also decreases like a power law by increasing size of samples and goes below $0.2\%$ with $n=3\times 10^4$ samples. Since the expected relative entropy is computationally expensive, the code provide a class to parallel all computations using the MPI. 

\begin{figure}
	\centering
	\includegraphics[width=.5\textwidth]{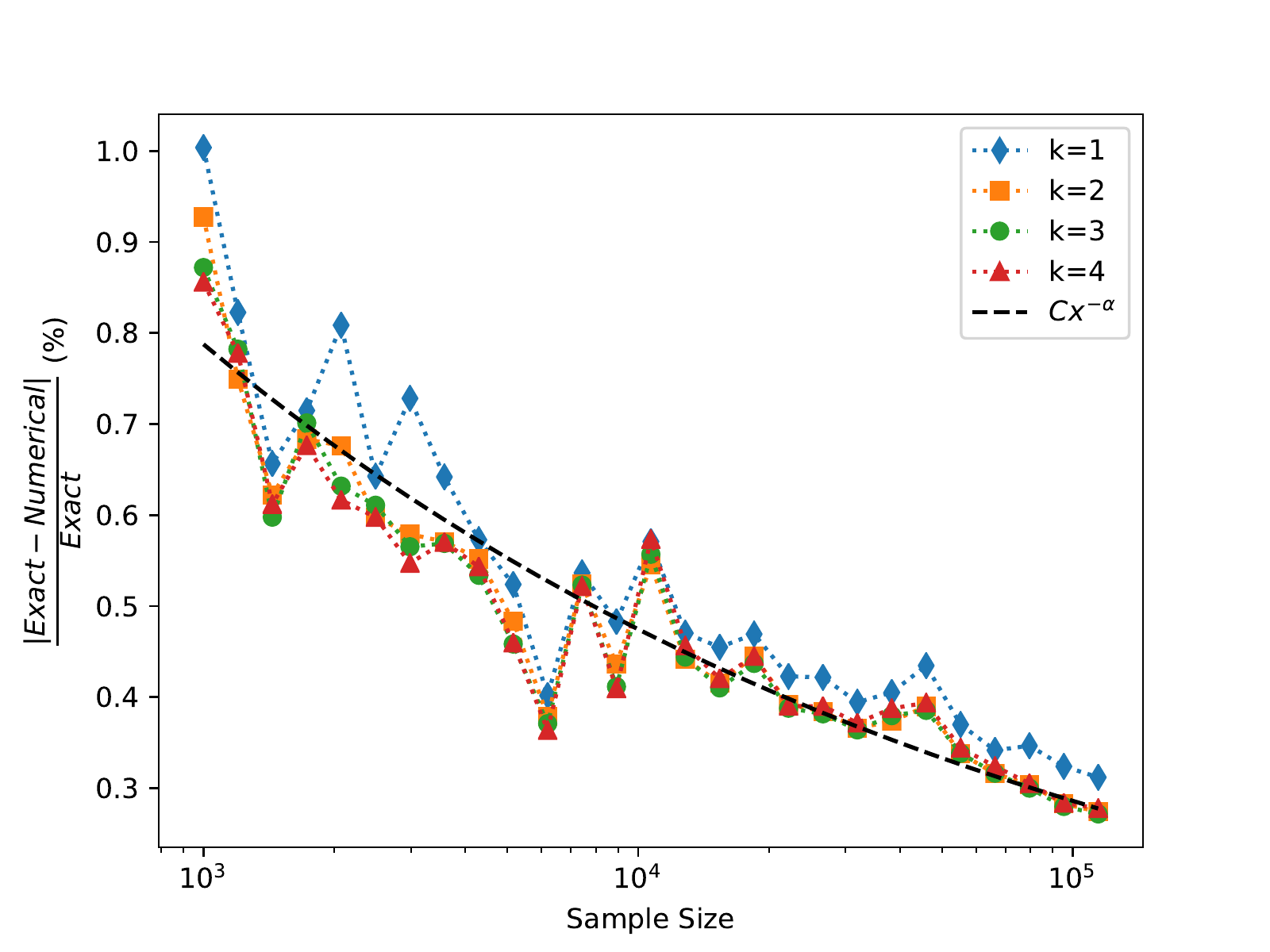}
	\caption{Relative difference of the exact and estimated relative entropy in 5D case.}
	\label{fig:rel-ent-5d}
\end{figure}

\begin{figure}
	\centering
	\includegraphics[width=.5\textwidth]{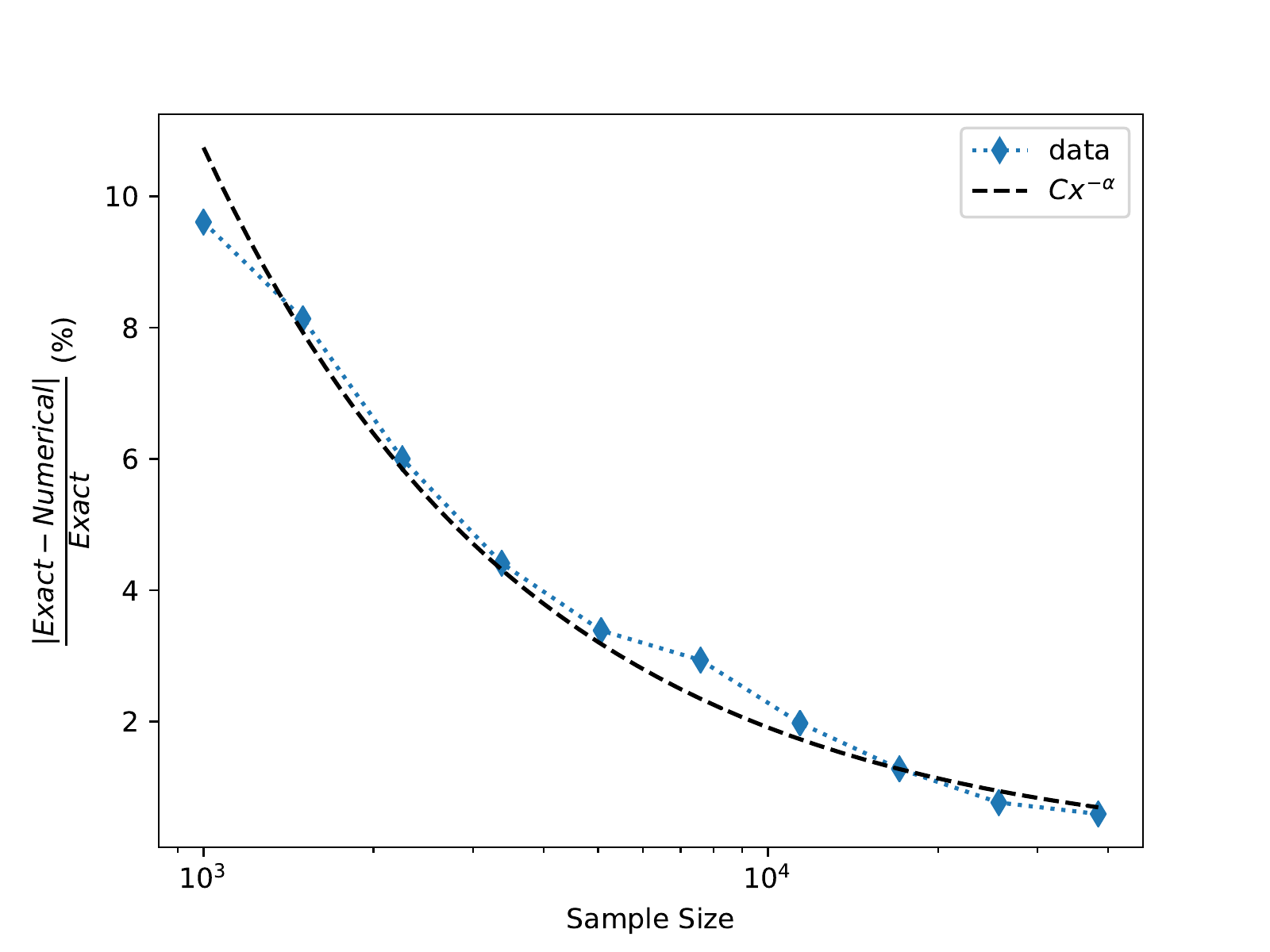}
	\caption{Relative difference of the exact and estimated expected relative entropy in 5D case.}
	\label{fig:rel-eent-5d}
\end{figure}

Moreover, we repeat above computation for a 5 dimensions case to study how increasing dimension affects the results. The results in this case, are presented in  Figs (\ref{fig:rel-ent-5d} and \ref{fig:rel-eent-5d})  

In contrast to 3D, different value of k gives almost the same results. In this case the relative error for $10^5$ samples is around $0.3\%$ which indicates robustness of our code. Similar to the 3D case, the expected relative entropy decreases like a power law and is around $0.5\%$ for $3\times10^4$ samples. Note that, error in the expected relative  entropy for 5D case is relatively larger than 3D with small number of samples.  

\section{Conclusion}\label{conclude}
In this work, we introduce a novel method and provide its python package to compute relative and expected relative entropy.  The relative entropy quantifies amount of information in updating from a prior to a posterior probability densities. Since in the most cases of Bayesian inference, we use an MCMC algorithm to generate a sample of posterior, our code use the chain information alongside the $\log(\rm prior)$ to estimate the relative entropy from the chain. For expected relative entropy, the relative error between the exact and estimated value in a linear Gaussian model are around $0.2\%$ and $0.5\%$ for $10^5$ samples in 3D and 5D respectively.  Since there is no closed-form solution for the relative entropy in the case of an arbitrary probability distributions, The code would be useful to estimate amount of information gain in updating from a prior to posterior in a general case. 

In addition to the relative entropy, there are some algorithms to estimate the expected relative entropy. The expected relative entropy has been used to define a quantity so called surprise. The surprise is a measurement of consistency between  posterior distributions and can be used to quantify possible tension between data sets within a model. Given a sample of first posterior and likelihood function of the second data set, our code provides an estimation of expected relative entropy base on the algorithm presented in  \cite{Huan:2013}.  Since the linear Gaussian model has an analytic solution, we compare  the estimated value with the exact one in 3D and 5D to check the robustness of our code. The relative errors in this case are around  $0.2\%$ and $0.5\%$ for $3\times10^4$ samples in 3D and 5D respectively. The code is available in \href{https://github.com/ahmadiphy/MCKLdivergence}{Github}. Alongside the code, there are two examples for computing the relative entropy and the expected relative entropy in the case of linear Gaussian model.


 \bibliographystyle{apsrev4-1}
  \bibliography{ref}

\end{document}